\documentclass[10pt,journal,twocolumn]{IEEEtran}
\ifCLASSINFOpdf
   \usepackage[pdftex]{graphicx}
   \DeclareGraphicsExtensions{.pdf,.jpeg,.png}
\else
   \usepackage[dvips]{graphicx}
   \DeclareGraphicsExtensions{.eps}
\fi
\usepackage{amsmath}
\usepackage{amssymb}
\usepackage{amsthm}
\usepackage{cancel}
\usepackage{breqn}
\usepackage{multirow}
\usepackage{rotating}
\usepackage{cite}
\usepackage{verbatim}
\DeclareMathOperator*{\argmax}{arg\,max}
\DeclareMathOperator*{\argmin}{arg\,min}
\DeclareMathOperator*{\sign}{sign}
\DeclareMathOperator*{\erfc}{erfc}
\newcommand{\norm}[1]{\left\lVert#1\right\rVert}
\newcommand{\ebn}{\ensuremath{E_{\mathrm{b}}/N_{\mathrm{0}}\ }}
\usepackage{esdiff}
\newtheorem{proposition}{Proposition}
\begin{document}
\title{Dual Polarized Modulation and Reception for Next Generation Mobile Satellite Communications}
\author{Pol~Henarejos,~\IEEEmembership{Student,~IEEE,}, %
        ~Ana~Perez-Neira,~\IEEEmembership{Senior,~IEEE}% <-this % stops a space
\thanks{P. Henarejos is with the Communications Systems Division of Catalonian Technological Telecommunications Center, Barcelona, e-mail: pol.henarejos@cttc.es.}% <-this % stops a space
\thanks{A. Perez-Neira is with the Signal and Communication Theory Department of Universitat Politècnica de Catalunya and with the Catalonian Technological Telecommunications Center, e-mail: ana.isabel.perez@upc.edu}%
\thanks{Manuscript submitted on October 15th 2014}}
%\author{Authors}

\markboth{Submitted to IEEE Transactions on Communications on October 15th 2014}%
{Author \MakeLowercase{\textit{et al.}}}%



\maketitle

\begin{abstract}
This paper presents the novel application of Polarized Modulation (PMod) for increasing the throughput in mobile satellite transmissions. One of the major drawbacks in mobile satellite communications is the fact that the power budget is often restrictive, making unaffordable to improve the spectral efficiency without an increment of transmitted power. By using dual polarized antennas in the transmitter and receiver, the PMod technique achieves an improvement in throughput of up to $100$\% with respect to existing deployments, with an increase of less than $1$ dB at low \ebn regime. Additionally, the proposed scheme implies minimum hardware modifications with respect to the existing dual polarized systems and does not require additional channel state information at the transmitter; thus it can be used in current deployments. Demodulation (i.e. detection and decoding) alternatives, with different processing complexity and performance, are studied. The results are validated in a typical mobile interactive scenario, the newest version of TS 102 744 standard (Broadband Global Area Network (BGAN)), which aims to provide interactive mobile satellite communications. 
\end{abstract}


% \fi
\IEEEpeerreviewmaketitle

\section{Introduction}
\IEEEPARstart{M}{ultiple}-input multiple-output (MIMO) schemes were introduced as a promising way to notably increase the spectral efficiency (SE) using multiples antennas at transmission and/or reception~\cite{SatelliteCommun.2008,Schwarz2008,Electr.amp2011}. In contrast to terrestrial communications, where the transmitter can obtain Channel State Information at the Transmitter (CSIT), in satellite links it is impossible to maintain CSIT updated due to the long round trip time. By the time the satellite receives the feedback parameters, the channel varies and the CSIT becomes outdated.

Among the different approaches that do not use CSIT, Vertical Bell Laboratories Layered Space-Time (VBLAST) scheme and successive improvements present a simple way to increase the achievable rate with a relative increase of the processing complexity~\cite{Wolniansky1998,Golden1999,Shen2003,Xue2003} in the absence of CSIT. However, VBLAST introduces interference between the streams since all signals are transmitted through all antennas without any interference pre-cancellation. In consequence, the signals must be transmitted with higher amplitude to obtain the same error rates compared with the single stream case.

In contrast to VBLAST, Spatial Modulation (SM) appeared recently to increase the SE~\cite{Mesleh2008,Chau2001,Mesleh2006}. In SM, the bits of information are split in blocks; some are coded as antenna indices and the rest are transmitted through the antennas that are selected by the indices. By estimating the antenna indices, the receiver can recover the bits used for this purpose. Nevertheless, this approach is very sensitive to the channel variations and requires an accurate channel estimation as well as spatially uncorrelated channels~\cite{SignauxetSyst.2012,DiRenzo2011,Mesleh2007}.% The receiver can enhance the detection if the channel's correlation is low and exploit the fact the signal can be received from individual paths. Thus, although the terminal is moving, it may be able to detect from which antenna the information is being conveyed.

In satellite scenarios, due to the Line of Sight (LOS), the spatial components become correlated at the receiver side although the transmitting antennas may be separated at half wavelength. Hence, in the absence of scatterers, the receiver only discovers a single transmission path and the sensitivity of the terminal is not enough to distinguish the different spatial signatures and detect the antenna indices. Because of this, SM does not seem suitable as it does not provide sufficient diversity in satellite scenarios.

On the contrary, the polarization channel provides more diversity and may be used for this kind of schemes~\cite{Horvath2006}. Although dual polarized antennas were used for broadcasting, where subscribers only tuned a single polarization, recent studies unveil that dual-polarized MIMO channel is richer in terms of diversity~\cite{&Technol.Centre2010}. Additionally, the use of dual polarized antennas is increasingly motivated by the new possibilities arising, together with the newest standards including dual polarized MIMO, such as Digital Video Broadcasting-Next Generation broadcasting system to Hand-held (DVB-NGH~\cite{dvb_ngh}). Finally, research projects such as~\cite{Henarejos,Gallinaro2014} reported that the throughput can be increased as in a conventional MIMO system if more antennas, and the consequent radio frequency (RF) chains~\cite{Vasseur2000,Electr.&Comput.Eng.2010,Tecnol.deTelecomunicacionsdeCatalunya2008}, are added in order to multiplex polarizations. The price to be paid is that the complexity of the satellite payload increases since interference among polarizations appears. For instance, extending the VBLAST strategy to dual polarized schemes requires higher transmit power to maintain the same Quality of Service (QoS) in point-to-point clients~\cite{Henarejos}. 

The primary motivation behind the present work is to apply the SM concept to dual polarized communications in the challenging mobile satellite channels. 
Hereafter, this scheme is referred to as polarized modulation (PMod). Indeed, it is not polarization multiplexing since only one polarization is activated at a time and therefore precludes the presence of interference. Although this work has been triggered as an attempt to apply a simple diversity technique as SM to the satellite scenario, the paper could also be viewed as the extension to satellite communications of the PMod idea that has been reported previously in optical communications~\cite{Karlsson:09}. However, from the authors' knowledge there is no literature describing in detail PMod demodulation scheme (detection and decoding) for optical communications, being polarization multiplexing far more common.

The scheme to be proposed contains the following contributions:
\begin{itemize}
\item The proposed PMod technique exploits the polarization diversity in satellite scenarios, where the spatial diversity is highly penalized.
\item PMod does not require CSIT and increases the throughput maintaining the robustness based on the polarization diversity.
\item Usually, satellite systems operate with dual polarized antennas and thus the proposed scheme does not require additional antennas. The minimum requirement is to use a dual polarized feeder.
\item The success of this scheme lies not only on the simplicity of the transmission technique, but also on the receiver design, which is also one main contribution of the present work, together with the performance evaluation. Note that the information is conveyed not only in the transmitted bit stream, but also in the polarization.
\item Finally, as we demonstrate in a maritime mobile satellite L-band scenario, the result is an increase of the overall performance in terms of throughput, whereas guaranteeing a minimum QoS and requiring a minimum increase in power usage. The best performance is obtained for low order modulations, where the proposed method achieves a gain of $2$ when it is compared with a basic system without PMod. 
\end{itemize}

\section{System Model}
Let us consider a MIMO system where transmitter and receiver are equipped with a single antenna with dual polarization, and a Rician frequency flat channel. Each symbol contains $b+1$ bits of information, where $b$ bits are modulated within the constellation $\mathcal{S}$ and the remaining bit is used for polarization selection. This remaining bit is denoted as $c$ and the modulated bits as $s\in\mathbb{C}$. We would like to remark that the information is conveyed through the symbol $s$ as well as bit $c$. The channels across the polarizations $1$ and $2$ are denoted $h_{11}\in\mathbb{C}$ and $h_{22}\in\mathbb{C}$, respectively, and their respective cross-channels as $h_{21}\in\mathbb{C}$ and $h_{12}\in\mathbb{C}$. All channel coefficients $h_{ij}$ follow a Rice statistical distribution with ($K$,$\sigma_h$) parameters with a pair-wise correlation $\rho_{ij}=[0,1]$. The received signals for polarizations $1$ and $2$ are denoted as $y_1\in\mathbb{C}$ and $y_2\in\mathbb{C}$, respectively, and $\omega_i\in\mathbb{C}$ follows Additive White Gaussian Noise (AWGN), $\boldsymbol\omega\,\mathtt{\sim}\,\mathcal{CN}\left(\mathbf{0},N_0\mathbf{I}_2\right)$. 

Depending on the value of $c\in\{0,1\}$, the $s$ symbols are conveyed using one polarization or the other. Hence, we can formulate the system model as follows:
\begin{equation}
\left[\begin{array}{c}y_1\\y_2\end{array}\right] = \left[\begin{array}{cc}h_{11} & h_{21}\\h_{12} & h_{22}\end{array}\right]\left[\begin{array}{c}1-c\\c\end{array}\right]s+\left[\begin{array}{c}\omega_1\\\omega_2\end{array}\right] 
\label{eq:sys_mod}
\end{equation}
or in a more compact way as:
\begin{eqnarray}
\mathbf{y} &=& \left[\mathbf{h}_1 \mathbf{h}_2\right] \mathbf{c}s+\boldsymbol\omega\\
&=&\mathbf{Hc}s+\boldsymbol\omega,
\end{eqnarray}
where $\mathbf{h}_i$ is the channel corresponding to the $i$th polarization.

Since this scheme adds an additional bit to the transmission by keeping the same power budget, the throughput gain of PMod with respect to Single-input Single-output (SISO) case is
\begin{equation}
G=\frac{b+1}{b}=1+\frac{1}{b}.
\label{eq:G}
\end{equation}

For higher order modulations,~(\ref{eq:G}) is asymptotically $1$ and thus the proposed PMod scheme increases the gain for low order modulations. For instance, the gain is $2$ for Binary Phase-Shift Keying (BPSK) modulation or $1.5$ for Quadrature Phase-Shift Keying (QPSK) modulation. As low order modulations are used in low signal to noise ratio (SNR) regime, it is clear that PMod increases significantly the throughput gain $G$ in low SNR systems. This is exactly the scenario for mobile satellite communications where shadowing, fading and power limitations cause low SNR.

\section{Demodulation Schemes} 
The implementation of the receiver derives into several approaches depending on the scenario constraints. Since PMod transmits a single stream, we aim to extract this stream to be processed into a SISO decoder. This scheme offers two main advantages:
\begin{itemize}
\item Reduces the complexity drastically since the signal processing is one dimensional.
\item Can be combined with existing SISO decoders, maintaining the compatibility with the current standards.
\end{itemize}

\begin{figure}[!ht]
  \centering
    \includegraphics[width=0.7\linewidth]{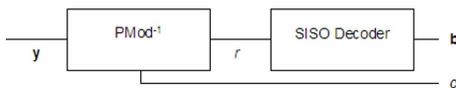}
    \caption{Reception scheme. PMod$^{-1}$ applies one of the following demodulation schemes to estimate the bit $c$ and prepare the signal $r$ to be processed by a common SISO decoder.}
    \label{fig:demod_scheme}
\end{figure}

The reception scheme is illustrated in Fig.~\ref{fig:demod_scheme}.

With this scheme, PMod$^{-1}$ implements one of the four demodulation schemes that are introduced in this section, estimates the bit $c$ and manipulates the received signal $\mathbf{y}$ to produce the signal $r$, which is capable to be used by a SISO decoder.

The optimal demodulation scheme is based on the Maximum a Posteriori (MAP) criteria, which is equivalent to the Maximum-Likelihood (ML) criteria for the case where the transmitted symbols are equiprobable. Thus, the optimal receiver can be derived from the expression
\begin{equation}
\hat{\mathbf{x}}=\argmax_{\mathbf{x}\in\mathcal{M}}\mathbf{p}\left(\mathbf{y}|\mathbf{x},\mathbf{H}\right)
\label{eq:ml1}
\end{equation}
where $\mathbf{x}=\mathbf{c}s$ and $\mathcal{M}$ is the symbol alphabet of $\mathbf{x}$. Hereafter, we assume that the noise contribution is Gaussian. Note that there is no restriction with the characteristics of the channel matrix. Thus, we do not take any assumption on the statistical independence of $\mathbf{H}$.

Hence,~(\ref{eq:ml1}) can be simplified as
\begin{equation}
\hat{\mathbf{x}}=\argmin_{\mathbf{x}\in\mathcal{M}}\norm{\mathbf{y}-\mathbf{Hx}}^2.
\label{eq:ml2}
\end{equation}

Based on the expression proposed in~(\ref{eq:ml2}) four different demodulators are proposed:
\begin{itemize}
\item First approach: zero forcer.
\item Second approach: per-symbol detection.
\item Third approach: per-hard-bit detection.
\item Fourth approach: per-soft-bit detection.
\end{itemize}

\subsection{First Approach: Zero Forcer}
\label{sec:a1}
In this approach a zero forcing pre-filter is applied before detecting the information in $\mathbf{x}$. We note that the zero forcer is the solution of $\diff{f}{\mathbf{x}^H}=0$, where $f(\mathbf{x})=\norm{\mathbf{y}-\mathbf{Hx}}^2$. This is equivalent to apply the filter
\begin{equation}
\mathbf{W}=\left(\mathbf{H}^H\mathbf{H}\right)^{-1}\mathbf{H}^H.
\end{equation}

Therefore we obtain the signal after the processing as follows:
\begin{equation}
\mathbf{z}=\mathbf{W}\mathbf{y}=\mathbf{c}s+\left(\mathbf{H}^H\mathbf{H}\right)^{-1}\mathbf{H}^H\boldsymbol\omega.
\end{equation}

In the case where the signal is being transmitted through polarization $1$, i.e. $c=0$, $z_1$ contains the signal plus noise and $z_2$ only receives the noise; and the reciprocal case for $c=1$, $z_1$ contains only the noise and $z_2$ conveys the signal plus the noise.  Therefore, to decide on $c$ we propose a power detector. It is denoted as:
\begin{equation}
\hat{c}=\argmax_i\left(\left|z_i\right|^2\right)-1.
\label{eq:dpow}
\end{equation}
Once the receiver knows the used polarization path, it is able to decode the symbol $s$ based on the signal $r=z_{\hat{c}+1}$. 

This solution presents a simple implementation but it is not a sufficient statistic to decode the whole problem (symbol $s$ and bit $c$). Since it computes the envelope of the vector, the information conveyed through the phases is lost and therefore it is not sufficient.

Nevertheless, this solution presents a major disadvantage. $\mathbf{H}^H\mathbf{H}$ could be badly conditioned and thus it may produce an excessive noise enhancement, doing impossible the demodulation.

\subsection{Second Approach: Per-Symbol Detection}
\label{sec:a4}
As we stated in the previous section, applying the $\mathbf{W}$ filter may introduce important distortions. Since the solution have to lie in the subset $\mathcal{M}$, the solution of the first approach may not be the optimal. Thus, the optimal approach to solve~(\ref{eq:ml2}) is performing an exhaustive search over the subset $\mathcal{M}$.

In the particular case of PMod, the transmitted vector $\mathbf{x}$ can be restricted to the subset $\mathbf{x}\in\mathcal{M}$, where the first vector $\left[\begin{smallmatrix}s\\0\end{smallmatrix}\right]$ of the set defines the transmission using the first polarization and the second vector $\left[\begin{smallmatrix}0\\s\end{smallmatrix}\right]$ of the set, using the second polarization. Hence, the decision rule for demodulating the bit $c$ is based on $\hat{c}=0$ if $\hat{\mathbf{x}}=\left[\begin{smallmatrix}s\\0\end{smallmatrix}\right]$ and $\hat{c}=1$ if $\hat{\mathbf{x}}=\left[\begin{smallmatrix}0\\s\end{smallmatrix}\right]$. In the same way, the signal $r$ can be written as $r=x_{\hat{c}+1}$.

This scheme, however, presents a notable increase of computational complexity. The exhaustive search requires to find the solution among several possibilities. The complexity of the initial search is $ \mathcal{O}\left(2^{b^2}\right)$ but due to the restriction of the set aforementioned, the complexity is reduced to $\mathcal{O}\left(2^{b+1}\right)$, which is equivalent to the complexity of the SISO case where $b+1$ bits are conveyed.

Furthermore, the previous demodulation schemes introduce hard decisions that induce non-linearities, such as sign() or abs() functions. In the presence of coded information, as it can be seen in~\cite{Fertl2012}, soft decoding outperforms the previous ML implementation. In the following sections we describe schemes that introduce soft information.

\subsection{Third Approach: Likelihood Ratio with Hard Decision}
\label{sec:a2}
Usually, to deal with channel impairments, the transmitted bits are coded. The channel decoder computes the metrics based on the likelihood of the received signal and is able to estimate the uncoded bits. 

The third approach is based on this philosophy and bit $c$ is estimated based on the likelihood ratio. If the likelihood ratio is defined as
\begin{equation}
\Lambda\left(\mathbf{y}\right)=\frac{P_2}{P_1}=\frac{P\left(c=1|\mathbf{y}\right)}{P\left(c=0|\mathbf{y}\right)}=\frac{\sum\limits_{\tilde{s}\in\mathcal{S}}\exp\left(-\frac{\|\mathbf{y}-\mathbf{h}_2\tilde{s}\|^2}{\sigma^2_w}\right)}{\sum\limits_{\tilde{s}\in\mathcal{S}}\exp\left(-\frac{\|\mathbf{y}-\mathbf{h}_1\tilde{s}\|^2}{\sigma^2_w}\right)},
\label{eq:lr_def}
\end{equation}
the decision rule for estimating $c$ depends only on the sign of $\log\left(\Lambda\left(\mathbf{y}\right)\right)$. In the case where likelihood ratio is greater than $1$, it means that it is more probable that $c=1$ and vice-versa. Assuming that only $b$ bits are coded, an estimator of the uncoded $c$ can be stated as:
\begin{equation}
\hat{c}=\frac{1+\sign\left(\log\left(\Lambda\left(\mathbf{y}\right)\right)\right)}{2}.
\end{equation}
Once the receiver obtains the estimation of $c$, it knows which polarization is being used and thus it can recover the symbol $s$ using the signal $r=y_{\hat{c}+1}$.

Although this scheme uses soft information in the decoding of symbol $s$, the decision of bit $c$ is still hard. Thus, if this bit is also coded, the result is suboptimal. In the next section we describe how to obtain a soft version of bit $c$.

\subsection{Fourth Approach: Likelihood Ratio with Soft Decision}
\label{sec:a3}
The three approaches described above perform hard decision for the estimation of bit $c$. However, they can introduce errors if the system conveys coded information as it was mentioned. 
The soft version of bit $c$ corresponds to the log-likelihood, exactly as the bits $b$. That is $\hat{c}=\log\left(\Lambda\left(\mathbf{y}\right)\right)$. 

After that, the bit $c$ is soft and can be passed to the soft decoder. However, there is the problem of which polarization to choose for decoding. In the previous schemes, since the bit $c$ is hard, it is possible to process the received signal on the polarization indicated by $c$. In the present scheme it is not possible to decide which polarization conveys information. 

To solve this issue we compute the average received signal:
\begin{equation}
r=P_1y_1+P_2y_2.
\end{equation}

Using the likelihood ratio $\Lambda\left(\mathbf{y}\right)$ computed as in~(\ref{eq:lr_def}), and using 
\begin{equation}
P_2=P\left(c=1|\mathbf{y}\right)=1-P\left(c=0|\mathbf{y}\right)=1-P_1,
\end{equation}
we can rewrite 
\begin{equation}
P_2=P\left(c=1|\mathbf{y}\right)=\frac{\Lambda\left(\mathbf{y}\right)}{1+\Lambda\left(\mathbf{y}\right)}.
\end{equation}

Therefore, the receiver can recover the signal by weighting the received signals from both polarizations by $P_1=1-P_2$ and $P_2$, respectively. If we assume that the bit $c$ is transmitted with equal probability, the averaged received signal takes the following form:
\begin{equation}
r=\left(1-P_2\right)y_1+P_2y_2=\frac{1}{1+\Lambda\left(\mathbf{y}\right)}\left(y_1+y_2\Lambda\left(\mathbf{y}\right)\right).
\end{equation}
Finally, the combined signal $r$ is passed to the decoder in order to obtain the $b$ bits.

\section{Numerical Results for Uncoded BER}
In this section we analyse the results of the proposed schemes. To compare them, we deploy a system conveying QPSK symbols in addition to the switching bit $c$. For this purpose, we only examine the uncoded bit error rate (BER). The channel model used corresponds to the Rician maritime mobile channel model described in the experiment V in~\cite{Sellathurai2006} with a correlation factor of $\rho_{ij}$. All parameters are summarized in Table~\ref{tab:param}.

In all results the following labels are used:
\begin{enumerate}
\item \emph{Reference} denotes the reference scenario, i.e. the scenario where single polarization is used.
\item \emph{VBLAST} is the polarization multiplexing VBLAST coding scheme.
\item \emph{PMod ZF} is the first approach described in Section~\ref{sec:a1}.
\item \emph{PMod ML} is the second approach described Section~\ref{sec:a4}.
\item \emph{PMod HD} is the third approach described in Section~\ref{sec:a2}.
\item \emph{PMod SD} is the fourth approach described in Section~\ref{sec:a3}.
\item \emph{OSTBC} corresponds to the Orthogonal Space Time Block Codes applied to polarization instead of spatial diversity~\cite{Perez-Neira2008}.
\end{enumerate}

\begin{figure}[!ht]
  \centering
    \includegraphics[width=1\linewidth]{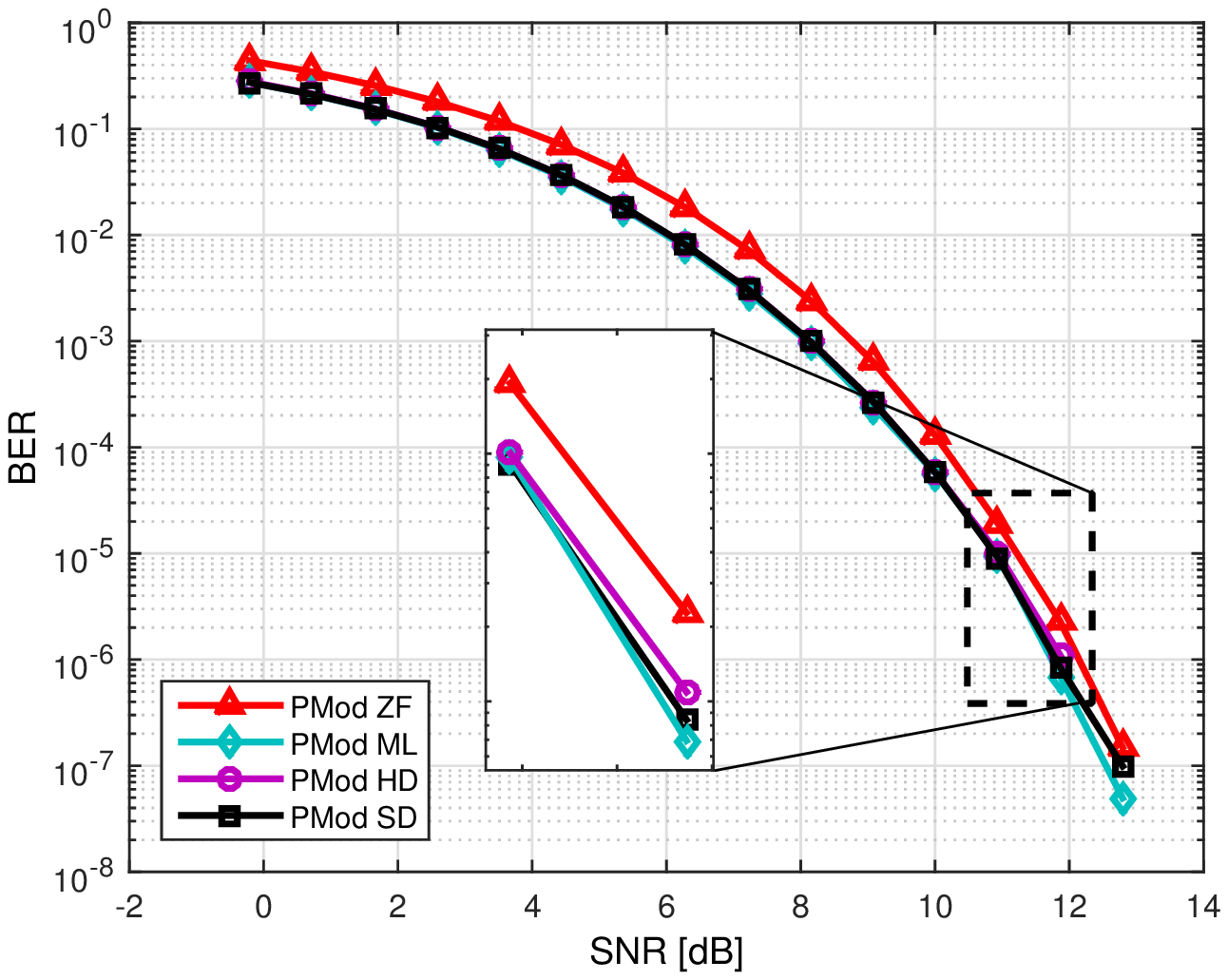}
    \caption{Comparison of the uncoded BER of the four proposed PMod techniques conveying a QPSK constellation.}
    \label{fig:BER_PMod}
\end{figure}

In Fig.~\ref{fig:BER_PMod}, we compare the BER of the four PMod schemes. The four curves are labelled in the same order that they have been introduced (from the first to the fourth approach).

As we stated in Section~\ref{sec:a4}, the ML solution provides the lowest error rate, immediately followed by the fourth solution. As expected, in the absence of channel coding, the ML receiver becomes the optimal solution. Although we remark that ML uses a reduced search space of order $\mathcal{O}\left(2^{b+1}\right)$, the computational complexity is sensibly higher with respect to the other solutions.

Next to the curve of ML is the pure soft scheme (the fourth). If we examine the magnified area, we observe the gap between the ML solution with the pure soft is tight. Hence, we conclude that the fourth demodulation scheme stays very close to the optimal solution.

Finally, the third approach, which does not use the conditional mean of the signal, performs close to PMod SD, whereas the first approach PMod ZF obtains the highest BER.

Hereinafter, we choose the fourth approach, PMod SD, to compare it with other schemes different to PMod. The reason is twofold:
\begin{itemize}
\item First, it performs a near-optimal ML solution, with a small gap of $0.05$ dB of \ebn for a fixed BER of $10^{-6}$.
\item Second, the computational complexity is less expensive than in ML.
\end{itemize}

\begin{figure}[!ht]
  \centering
    \includegraphics[width=1\linewidth]{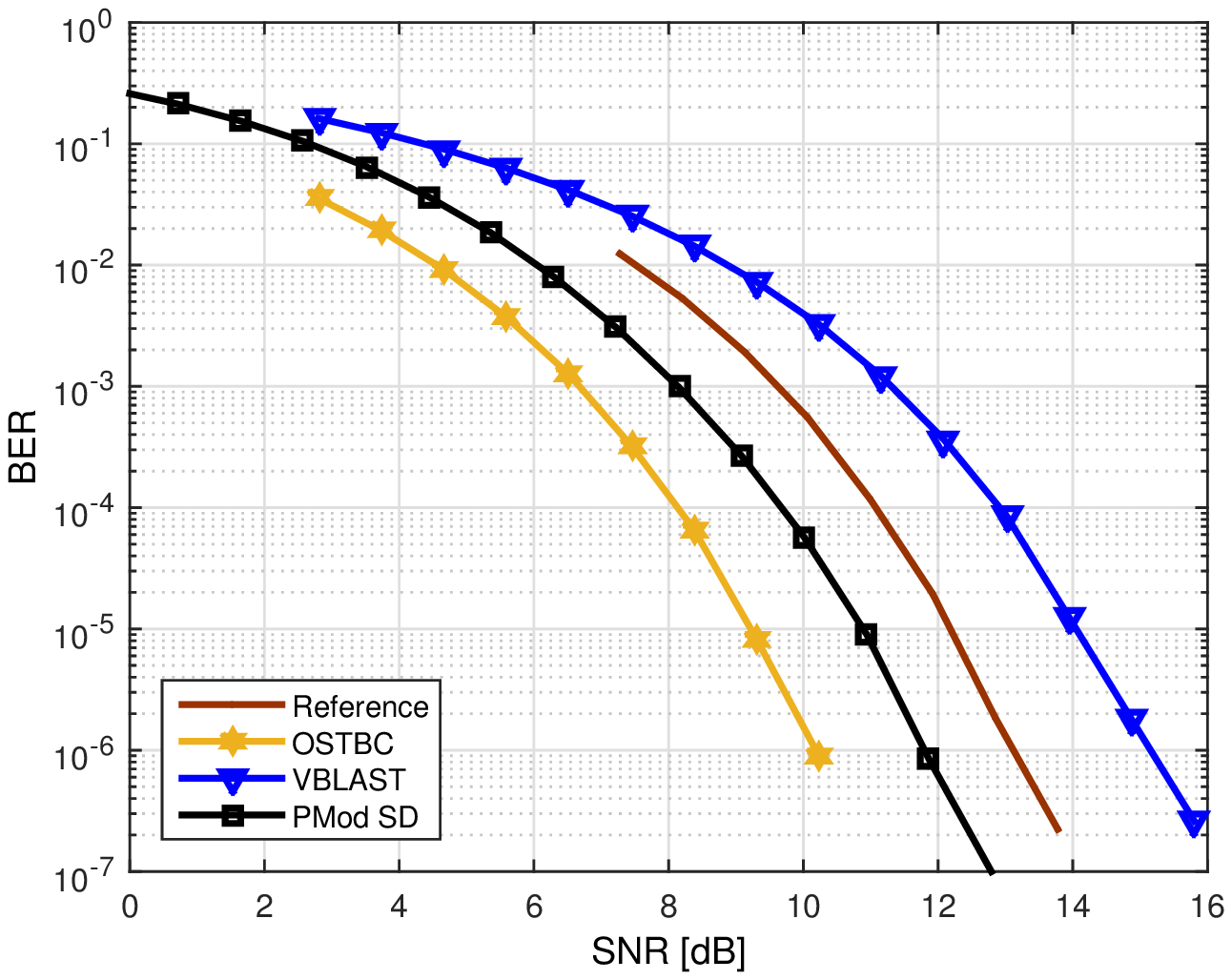}
    \caption{Comparison of the uncoded BER of the PMod SD with other existing techniques conveying a QPSK constellation.}
    \label{fig:BER_All}
\end{figure}

Fig.~\ref{fig:BER_All} compares the PMod SD solution with the conventional OSTBC, VBLAST and reference scenario using a QPSK constellation for all schemes. Note that even though we use the same constellation for all schemes, the total SE is different for each scheme. Thus, although we are comparing different schemes with different SE, the most remarkable point is the fact that the PMod is bounded by OSTBC (lower SE) and VBLAST (higher SE) and therefore PMod achieves a trade-off between OSTBC and VBLAST in terms of BER and SE.

In all these schemes, 2 bits/channel use are conveyed. As expected, OSTBC obtains lowest BER, followed by PMod and VBLAST. However, OSTBC does not allow to increase the granularity of the adaptive bit rate. In other words, there is no choice to transmit 3 bits/channel use. The next step is to transmit a 16QAM with OSTBC, which is 4 bits/channel use. Newest standards, such as DVB-S2X~\cite{dvb_s2x}, aim to include new modulation schemes to refine the rate adaptation curve.

\subsection{Equal SE Analysis}
In contrast to the previous section, where the comparison is performed maintaining the same constellation, in this section we analyse the performance of PMod compared with the other schemes but constraint to the same SE. To do this, we use the following transmission schemes:
\begin{itemize}
\item PMod SD with BPSK constellation.
\item VBLAST with BPSK constellation.
\item OSTBC with QPSK constellation.
\item Reference with QPSK constellation.
\end{itemize}

\begin{figure}[!ht]
  \centering
    \includegraphics[width=1\linewidth]{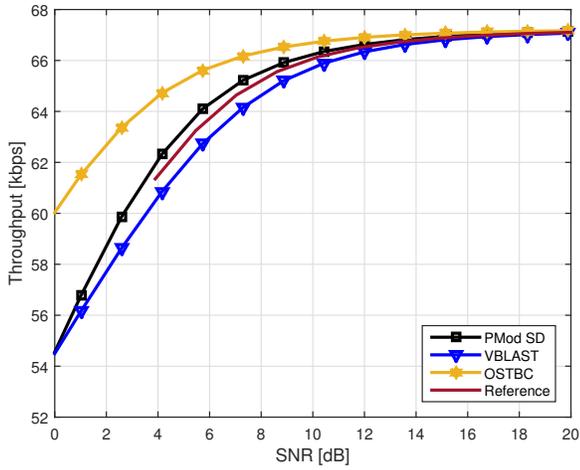}
    \caption{Comparison of the throughput of the PMod SD with other existing techniques constraint to the same SE.}
    \label{fig:throughput_eq}
\end{figure}

In all schemes, $2$ bits per channel use are conveyed. Fig.~\ref{fig:throughput_eq} describes the curves of the different throughputs and it is clear that all curves tend to the same throughput for high SNR.

\begin{figure}[!ht]
  \centering
    \includegraphics[width=1\linewidth]{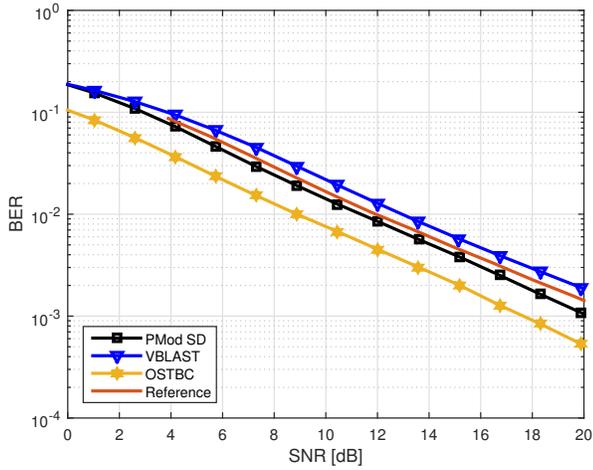}
    \caption{Comparison of the Uncoded BER of the PMod SD with other existing techniques constraint to the same SE.}
    \label{fig:ber_eq}
\end{figure}

Fig.~\ref{fig:ber_eq} depicts the BER for the different techniques. In this case, \emph{OSTBC} obtains the lowest BER, followed by \emph{PMod SD}, \emph{Reference} and \emph{VBLAST}, respectively. As expected, OSTBC exploits the full diversity of the channel and is closely followed by PMod. However, one of the advantages of PMod in front of OSTBC is the ability to increase the granularity of the throughput adaptation. Whereas OSTBC increases the throughput by a powers of two, PMod can increase the throughput by small fractions, as it is seen in~(\ref{eq:G}).

\section{Results in a Realistic System Context}
\label{sect:sim_bgan}
In this section we describe the implementation of the PMod solution inside the Broadband Global Area Network (BGAN) standard. In more detail, we deploy the downlink of the Next Generation Satellite Communications standard, currently being redacted at the European Telecommunications Standards Institute (ETSI) committee (more detail at~\cite{bgan}). This part of the standard defines the scrambling, turbo coding and mapping stages, among other procedures. In order to offer flexibility in terms of data rate, several bearers and subbearers are detailed. They are different profiles with many combinations of coding rate and constellations. Focusing in the downlink part, the symbol rate is $33.6$ ksps and the frame length is $80$ ms, where the blocks of coded symbols are not interleaved. In order to simplify the model, QPSK bearers will be used in all simulations.

\subsection{Next Generation Satellite Communications Simulation Framework}
\begin{figure}[!ht]
  \centering
    \includegraphics[width=1\linewidth]{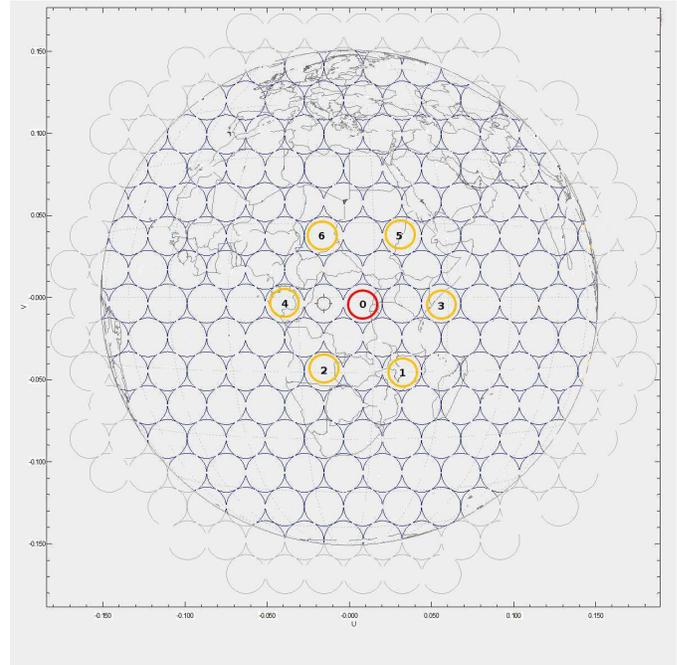}
    \caption{Considered beam pattern to perform realistic simulations. Working beam is marked with a red circumference and interfering beams as yellow circumferences.}
    \label{fig:beampatt}
\end{figure}

\begin{figure}[!ht]
  \centering
    \includegraphics[width=1\linewidth]{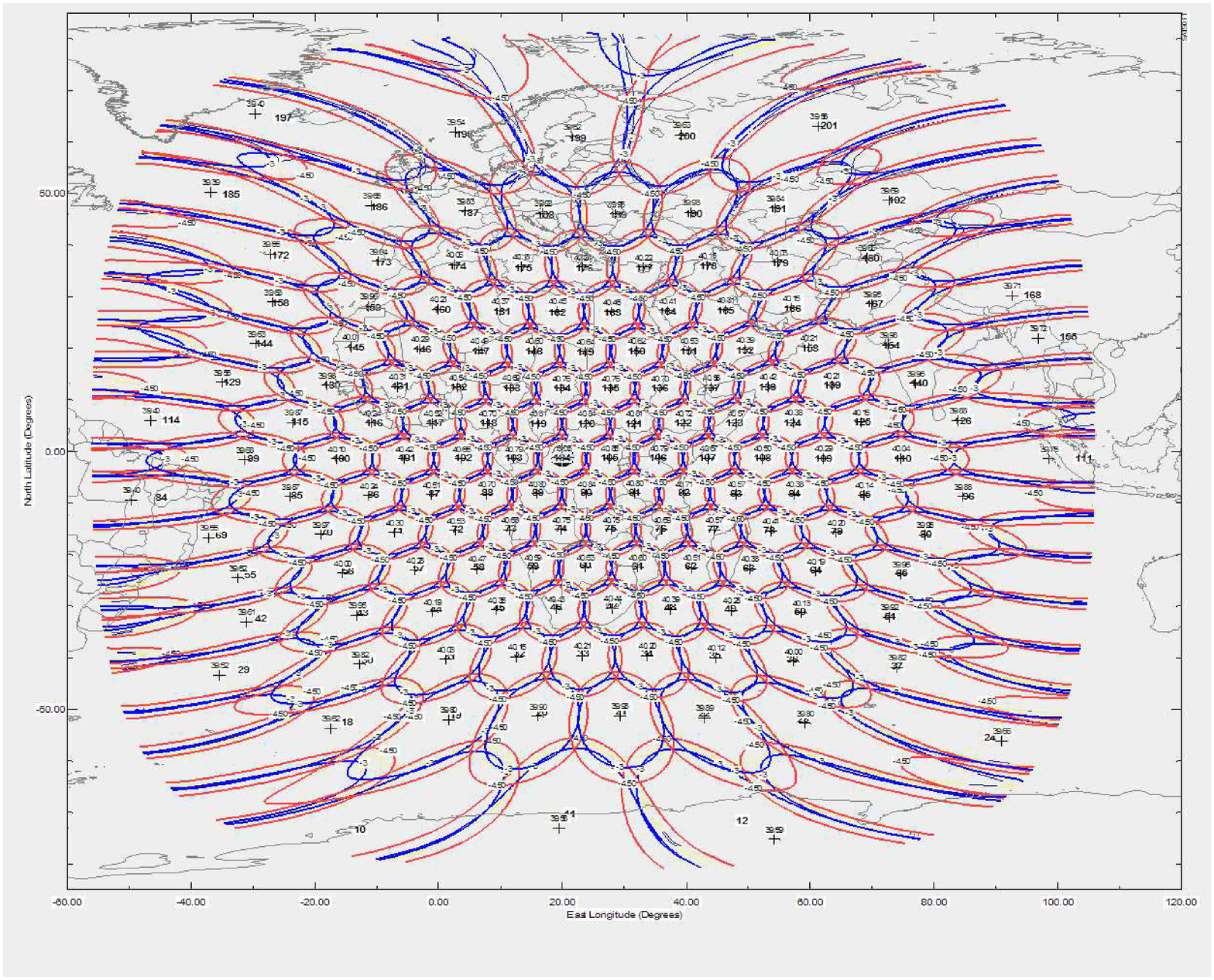}
    \caption{Co-polar coverage for the forward link with contours at $3$ dB (red lines) and $4.5$ dB (blue lines).}
    \label{fig:bp_cop}
\end{figure}

\begin{figure}[!ht]
  \centering
    \includegraphics[width=1\linewidth]{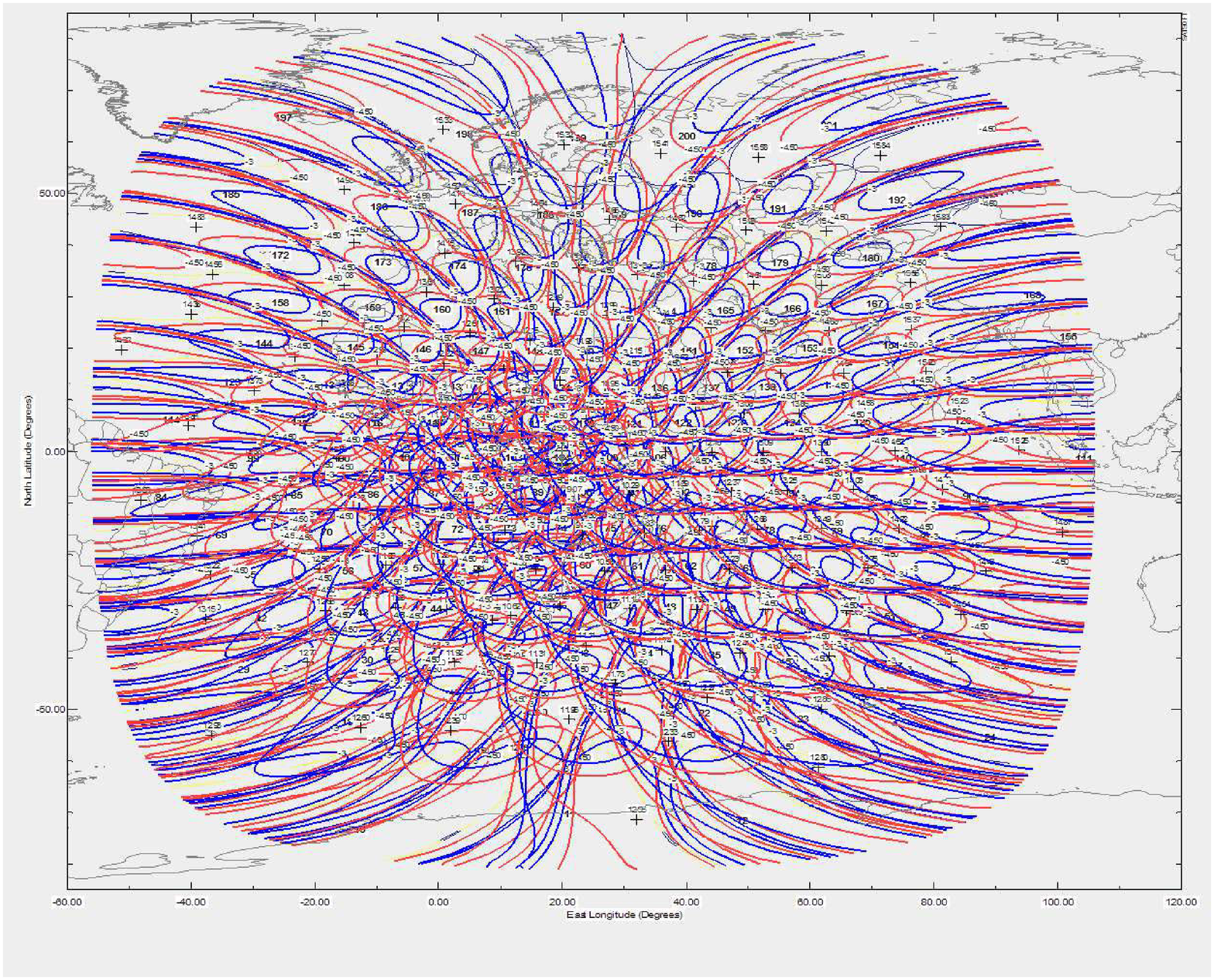}
    \caption{Cross-polar coverage for the forward link with contours at $3$ dB (red lines) and $4.5$ dB (blue lines).}
    \label{fig:bp_xpd}
\end{figure}

\begin{figure}[!ht]
  \centering
    \includegraphics[width=1\linewidth]{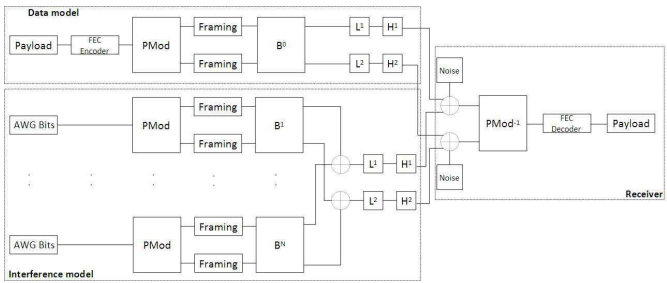}
    \caption{Block diagram of the simulation framework.}
    \label{fig:diagram}
\end{figure}

We simulate a L-band geostationary satellite with $7$ beams (one desired beam and six interfering beams) and dual polarization. Since the beams are not perfectly orthogonal, we consider six adjacent beams at the same frequency subband as interferences, as well as the cross polarization couplings. All these values are summarized in Table~\ref{tab:as1} and are obtained via realistic multibeam antenna pattern during the project Next Generation Waveforms for Improved Spectral Efficiency (NGW), whose results are summarized in~\cite{Henarejos}. In more detail, Fig.~\ref{fig:beampatt} illustrates the beam pattern, where the working beam is marked with a red circumference and the interfering beams as yellow circumferences. It is important to remark that not all beams induce the same levels of interference. Depending on the position of the satellite and the geometry of the reflectors, the power of interferences varies between beams. In more detail, Fig.~\ref{fig:bp_cop} and Fig.~\ref{fig:bp_xpd} illustrate the co-polar and cross-polar coverage for the forward link with contours at $3$ dB (red lines) and $4.5$ dB (blue lines). One of the relevant aspect is the asymmetry of the co-polar and cross-polar gains in each beam. From these figures, it is clear that gains are different for each beam spot. Finally, Fig.~\ref{fig:diagram} shows the block diagram used for the simulations described hereafter.

In Fig.~\ref{fig:diagram} the identified blocks are:
\begin{itemize}
\item Forward Error Correction (FEC) Encoder: encodes the bit stream using the specifications of~\cite{bgan}.
\item $PMod$: groups the bits in blocks of size $b+1$, modulates the symbols $s$ and uses the $c$ bits to select the polarization for each symbol.
\item Framing: encapsulates the symbols of each polarization in a frame defined in~\cite{bgan}. It inserts pilots for channel estimation, a preamble for synchronization and a header for modulation-code identification.
\item Interference matrix $\mathbf{B}_i$: models the cross polarization by a factor defined in Table~\ref{tab:as1}. $B_0$ corresponds to the cross-polarized matrix of intended data and $\mathbf{B}_1,\dots,\mathbf{B}_6$ correspond to the cross-polarized matrices of interfering beams.
\item $P$: the signal is amplified by a factor of $\sqrt{P}$. It is important to remark that this is possible due to the fact that, for each symbol, only a single polarization is active and thus all power budget $P$ is available, whereas in the case of \emph{VBLAST} and \emph{OSTBC} this factor is divided by $2$.
\item $L_i,\, i={0,1}$: equivalent path-loss for each polarization.
\item $H_i,\, i={0,1}$: convolves the signal using the Rician fast fading channel model.
\item Noise: adds the AWGN.
\item $PMod^{-1}$: implements one of the schemes.
\item FEC Decoder: performs the inverse operation of FEC Encoder. It implements a Turbo Coder with Systematic Recursive Convolutional Codes (SRCC).
\end{itemize}

We consider the Rician maritime mobile channel model described in the experiment V in~\cite{Sellathurai2006} and the parameters described in Table~\ref{tab:param}.

The aim is to evaluate the basic transmission and reception concepts and schemes; thus, in this work it is assumed perfect synchronization at the receiver side as well as perfect channel estimation. Prior to detection of symbol $s$, one of the four approaches is performed in order to estimate the bit $c$ and filter the received signal. 

We remark that this scenario includes nongaussian interference. Thus, as we described the PMod solution under this assumption, we need to cope with the interference to minimize it. To achieve that, the receiver implements a MMSE linear filter. This configuration mitigates the interferences from the other beams as well as the other polarization for the detection of symbol $s$. 

One important aspect is the Faraday Rotation (FR), which appears at L-band. This effect is caused by the free electrons in the ionosphere and causes a rotation of the polarization. Since it changes the polarization, FR may be critical in order to estimate the bit $c$. Fortunately, this effect can be reduced using a circular polarization or performing an estimation and assuming that the FR remains invariant during the time slot. An estimation of FR is described in \cite{Meyer2008} and it can be applied using the pilot symbols used by the channel estimation. Nevertheless, for the simulations, we assume that this effect is corrected.

Finally, in the next stage, the demodulated soft bits are passed to the turbo decoder and scrambled to obtain the information bits. In contrast to the previous section, since we consider interferences in this scenario, we use the signal to interference plus noise ratio (SINR) in the x-axis rather than SNR.

\subsection{Comparing PMod Solutions}
\begin{figure}[!ht]
  \centering
    \includegraphics[width=1\linewidth]{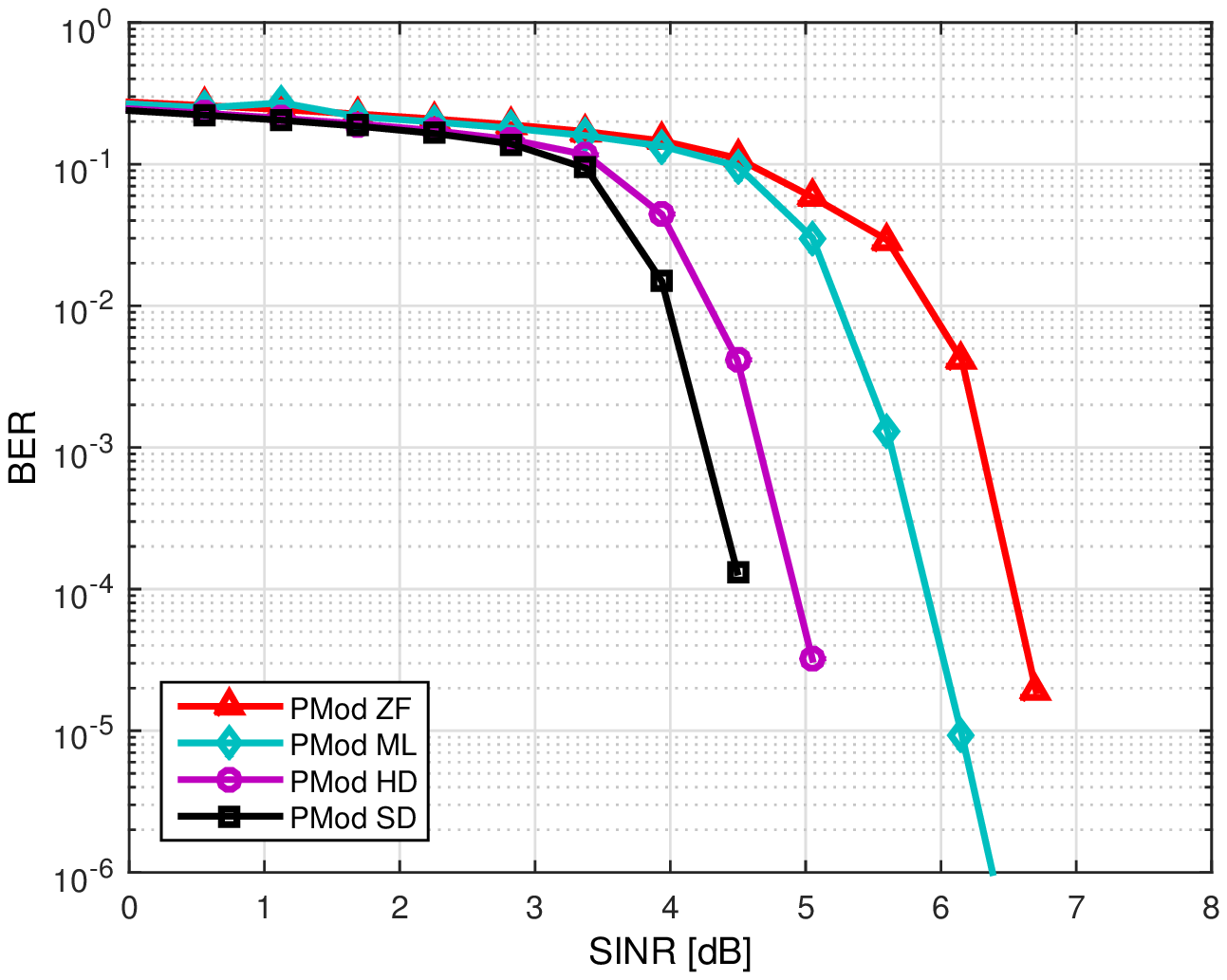}
    \caption{Comparison of the coded BER of the four proposed PMod techniques conveying a QPSK constellation.}
    \label{fig:ber_pmod}
\end{figure}

We compare the four proposed demodulation schemes. In contrast to Fig.~\ref{fig:ber_pmod}, although the ML solution is the optimal in absence of channel coding, this is not the case in the presence of coded information. Certainly, the PMod SD scheme produces the lowest BER, followed by PMod HD. Both schemes use soft bits and, thus, their performance is better than the hard solutions (PMod ZF and PMod ML). 

In order to compare the proposed schemes with the existing ones, we compare the performance in terms of throughput, which corresponds to the average rate of successful information delivery and is defined as
\begin{equation}
T=R(1-BLER)G.
\label{eq:thdef}
\end{equation}
This is equivalent to the bitrate ($R$) of the particular bearer weighted by the probability of no error in the whole block ($1-BLER$) and the throughput gain ($G$), defined in~(\ref{eq:G}). During all simulations, a fixed modulation-code is simulated with coding rate of $0.625$ ($R=40$ kbps without frame overhead). $BLER$ is obtained by simulations and corresponds to the number of erroneous blocks divided by the total number of blocks.

\begin{figure}[!ht]
  \centering
    \includegraphics[width=1\linewidth]{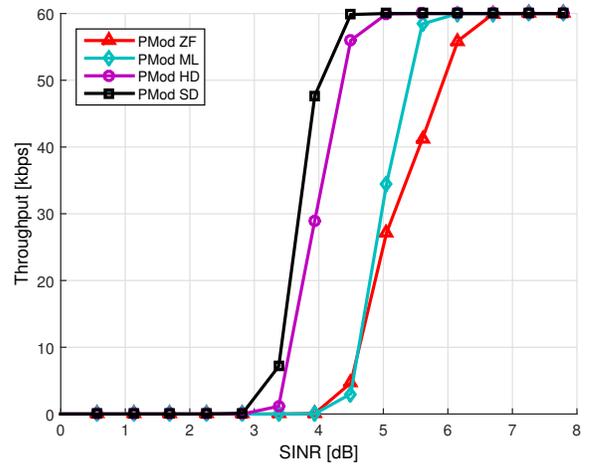}
    \caption{Comparison of the throughput of the four proposed PMod techniques conveying a QPSK constellation.}
    \label{fig:throughput_pmod}
\end{figure}

Fig.~\ref{fig:throughput_pmod} describes the throughput achieved using the four schemes. We observe that the four curves are grouped in the soft and hard receivers. In contrast to the previous section where the gap between the solutions is tight, in this case the gap increases notably, making clearer the performance of the PMod SD/HD. 

\subsection{Comparing PMod SD with Other Solutions}
In this section we compare the performance of the PMod SD with OSTBC and VBLAST in the same interference scenario. With the following comparisons we examine different strategies to increase the throughput.

\begin{figure}[!ht]
  \centering
    \includegraphics[width=1\linewidth]{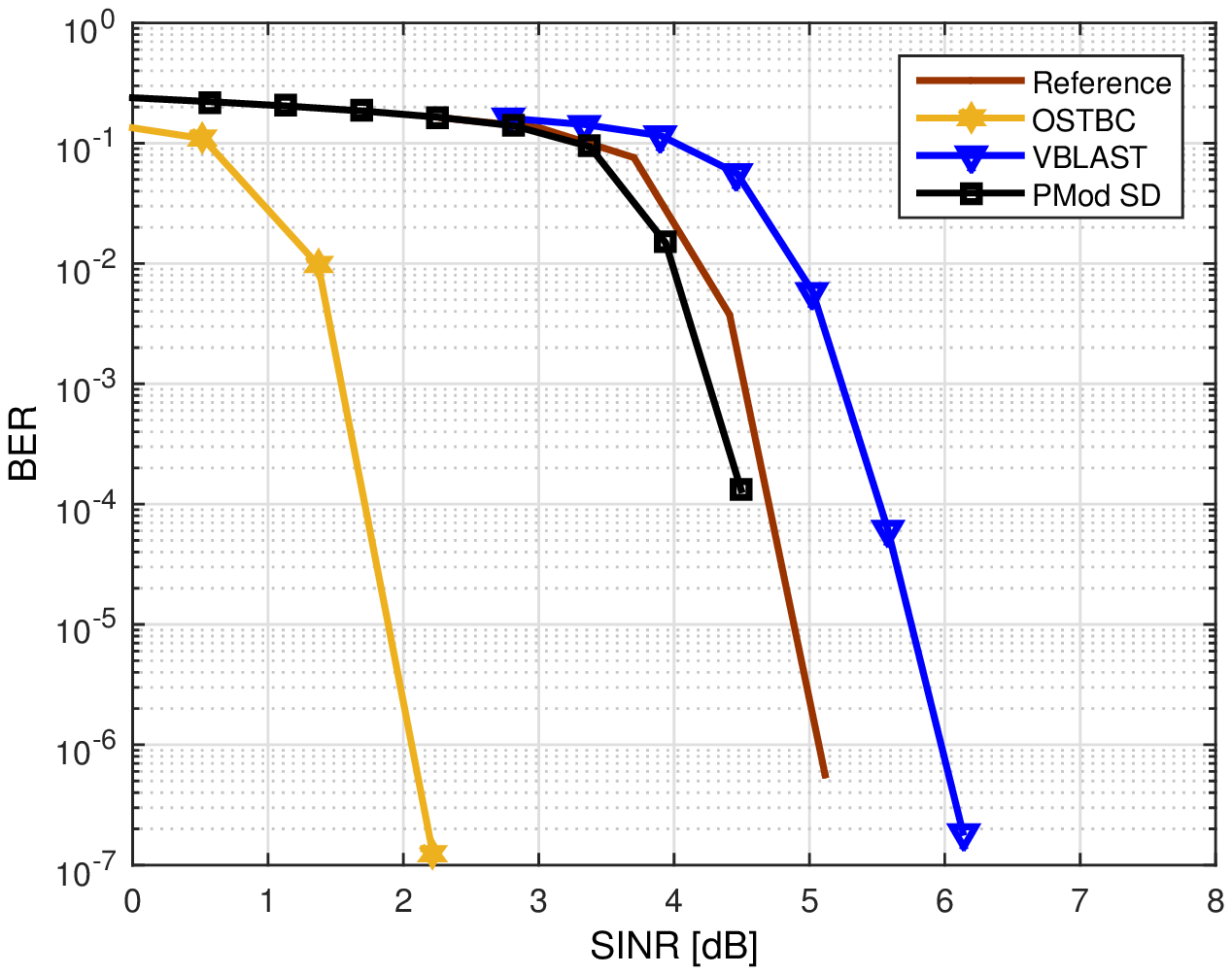}
    \caption{Comparison of the coded BER of the PMod SD with other existing techniques conveying a QPSK constellation.}
    \label{fig:ber}
\end{figure}

Fig.~\ref{fig:ber} illustrates the coded BER for the different schemes. As with the uncoded BER (see Fig.~\ref{fig:BER_All})), in this case, PMod SD lies between OSTBC and VBLAST. One important aspect is that improves the BER of the Reference scenario. This is positive since PMod increases the SE but also the error rate.

\begin{figure}[!ht]
  \centering
    \includegraphics[width=1\linewidth]{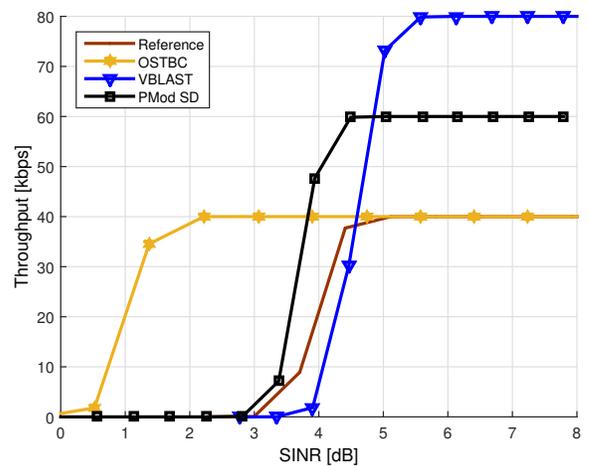}
    \caption{Comparison of the throughput of the PMod SD with other existing techniques conveying a QPSK constellation.}
    \label{fig:throughput}
\end{figure}

Finally, Fig.~\ref{fig:throughput} illustrates the throughput achieved by each scheme. The interesting part of this figure is the adaptation of the rate. For very low \ebn the most effective scheme is OSTBC. From $3.5$ dB, the PMod SD increases the throughput by a factor of $1.5$, followed by VBLAST from $5.5$ dB. This motivates the use of PMod in Adaptive Modulation and Coding Schemes (AMC).

\subsection{XPD Analysis}
In addition to prior comparisons, we also include a cross-polarization discrimination (XPD) analysis for the PMod. The results are extremely encouraging and reveal that the PMod scheme is robust in front of  cross-polarization impairments. The reason is twofold:
\begin{itemize}
\item For high XPD values, only one polarization carries the data symbol whereas the other only contains noise. In this case, the system will decode the symbols $s$ and the switching bits $c$ correctly.
\item For low XPD values, both polarizations carry the same symbol but only one polarization is decoded. In this case, the probability of error of decoding bit $c$ increases as the XPD decreases but the probability of error of decoding the symbol $s$ remains the same. This is motivated by the fact that, even in the case where the $c$ bit is erroneous and the decoded polarization is the wrong one, it also contains the symbol $s$ and thus, is able to decode the $s$ symbols as if it was decoded from the other polarization.
\end{itemize}

To analyse the XPD of the PMod technique, the XPD is defined as follows:
\begin{equation}
XPD=20\log\left(\frac{\left|y_c\right|}{\left|y_{1-c}\right|}\right),
\end{equation}
where $y_c$ is the signal received at the polarization where the symbol is transmitted and $y_{1-c}$ is the other one.

\begin{figure}[!ht]
  \centering
    \includegraphics[width=1\linewidth]{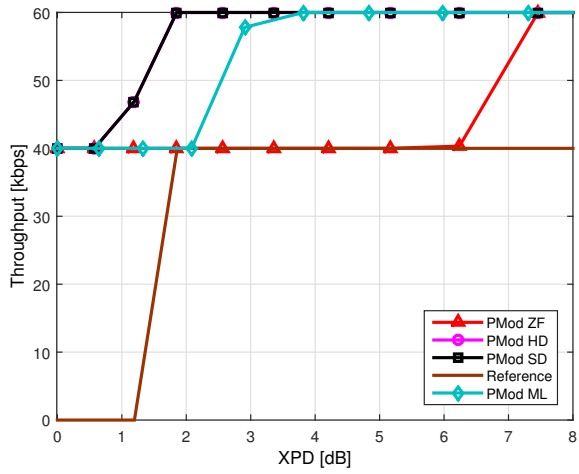}
    \caption{Comparison of the throughput with respect of XPD of the different techniques conveying a QPSK constellation.}
    \label{fig:xpd}
\end{figure}

Fig.~\ref{fig:xpd} compares the throughput of the four proposed schemes (\emph{PMod ZF}, \emph{PMod ML}, \emph{PMod HD} and \emph{PMod SD}) for different values to the XPD with the reference (\emph{Reference}). Note that \emph{PMod HD} and \emph{PMod SD} are overlapped in the fig., although the \emph{PMod SD} has slightly higher robustness. Particularly, only for these simulation a fixed SNR of $20$ dB was set, whilst the other parameters remain the same as in previous figures. As aforementioned, the PMod technique is robust in front of XPD as it can exploit the fact that the $2/3$ of the bits are transmitted through the both polarizations.

\subsection{Imperfect Channel Estimation}
\begin{figure}[!ht]
  \centering
    \includegraphics[width=1\linewidth]{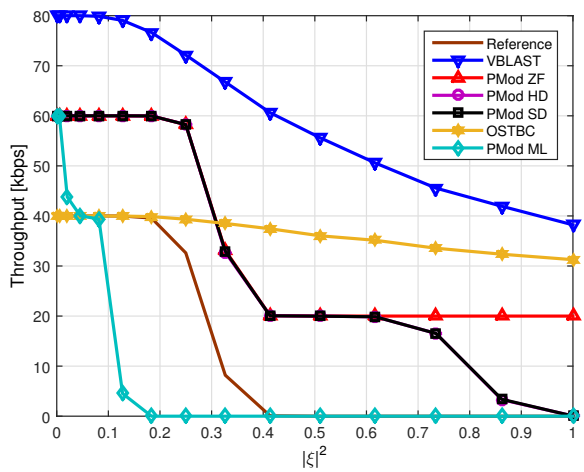}
    \caption{Impact of the imperfect channel estimation in the different techniques conveying a QPSK constellation.}
    \label{fig:impcsi}
\end{figure}

In this section we analyse the impact of an imperfect channel estimation. To study this behaviour we introduced an error $\xi$ into the channel estimation which is normalized by the channel norm. Indeed, the power of the error $\xi$ is defined as follows:
\begin{equation}
\left|\xi\right|^2=\frac{\mathbb{E}\left\{\left|h-\bar{h}\right|^2\right\}}{\mathbb{E}\left\{\left|h\right|^2\right\}}
\end{equation}
where $\mathbb{E}\left\{x\right\}$ is the expectation value of the variable $x$ and $\bar{h}$ is the estimated coefficient. The results can be examined in Fig.~\ref{fig:impcsi}.

The PMod scheme becomes more robust in front of the reference scheme (\emph{Reference}). Particularly, three of the four schemes (\emph{PMod ZF}, \emph{PMod HD} and \emph{PMod SD}) offer the same tolerance, but with the difference that \emph{PMod ZF} in addition is able to decode the bit $c$ correctly. This means that although the scheme may be inaccurate, it is always capable to decode the bit $c$. This motivates hierarchical modulations. For example, using the \emph{PMod ZF} we could establish a hierarchical BPSK+QPSK and always succeed on decoding the BPSK scheme at least.

Finally, it is worth mentioning that the PMod technique presents a good trade-off between robust techniques but with less throughput, such as OSTBC, and more throughput available techniques but more power consuming such as VBLAST, as Fig.~\ref{fig:impcsi} depicted.

\section{Conclusions}
This paper introduces a novel application to mobile satellite communications of the entitled Polarized Modulation, which is based on dual polarized antennas. The work shows that with dual-polarized modulation the throughput can be increased by a factor of $1+b^{-1}$ in the absence of CSIT in low \ebn regime and that the transmission results robust to cross-polarization and imperfect channel estimation. Performance depends on the implemented receiver, that is why in this paper different alternatives are proposed that trade-off computational complexity vs. performance. One of the demodulation schemes is based on probabilities, which involves soft detections and to the authors knowledge it is novel in the context of either spatial or polarized modulation. Finally, the proposed techniques have been thoroughly tested and validated using a maritime mobile satellite scenario and the newest implementation of the novel ETSI's standard TS 102 744~\cite{bgan}, known as BGAN, which is used for interactive mobile satellite communications. It validates the PMod scheme and demonstrates the enhancement of throughput and the robustness. Further work is to extend the results and receiver architectures to more than two polarizations and investigate the PMod in aeronautical and urban channels. PMod exploits the diversity of the channel and therefore, whereas the polarization channel has diversity, the PMod will work as expected. Additionally, although the union bound for Rayleigh channel is provided, an interesting action is to study the impact of the averaged probability of error in Rician channels as well as the mutual information and capacity analysis.

% or


\appendices
\section*{Acknowledgement}
The simulator tool was developed in the context of the European Space Agency (ESA) contract for the project Next Generation Waveform for Improved Spectral Efficiency and for the Satellite Network of Experts (SatNEx IV). The real measurements used as the interference gains as well as the coverage patterns were provided by ESA. 

This work has received funding from the Spanish Ministry of Economy and Competitiveness (Ministerio de Economia y Competitividad) under project TEC2014-59255-C3-1-R and from the Catalan Government (2014SGR1567).

We would also like to especially thank P. D. Arapoglou for his valuable comments and appreciations. We also kindly thank Prof. Miguel Angel Lagunas for his fruitful discussions.

\ifCLASSOPTIONcaptionsoff
  \newpage
\fi



\bibliographystyle{IEEEtran}
\bibliography{biblio}



%\vfill


\newpage

\begin{table}[!ht]
\centering
\caption{Scenario Main Parameters}
\begin{tabular}{|l|r|}
\hline
Profile & Maritime\\ \hline
Channel model & Rician flat fading\\ \hline
Rician K factor & 10\\ \hline
Doppler shift & $2$ Hz\\ \hline
Doppler spectrum & Jakes\\ \hline
Stream correlation & $\rho_{ij}=0.5$\\ \hline
Path distance & $35786$ km\\ \hline
Path loss & $187.05$ dB\\ \hline
Bandwidth & $200$ kHz\\ \hline
Terminal G/T & $-12.5$ dB/K\\ \hline
Carrier band & L ($1.59$ GHz)\\ \hline
Code rate & $0.625$\\ \hline
Bitrate & $40$ kbps\\ \hline
\end{tabular}
\label{tab:param}
\end{table}

\begin{table}[!ht]
\centering
\caption{Data Coupling Polarization Matrix + Interference Matrices}
\begin{tabular}{c|c|c|}
\cline{2-3}
 & Index & Interference matrix (dB)  \\
\cline{1-3}
\multicolumn{1}{|c|}{Data} & $0$ & $\left(\begin{array}{cc}
40.8 & -11.6 \\ -11.6 & 40.8
\end{array} \right)$  \\
\hline
\hline
\multicolumn{1}{|c|}{} &
$1$ & $\left(\begin{array}{cc}
3.7 & -12.3 \\ -12.3 & 3.7
\end{array} \right)$  \\
\cline{2-3}
\multicolumn{1}{|c|}{} & $2$ & $\left(\begin{array}{cc}
8.7 & -13\\ -13 & 8.7
\end{array} \right)$ \\
\cline{2-3}
\multicolumn{1}{|c|}{} & $3$ & $\left(\begin{array}{cc}
3.6 & -6.7 \\ -6.7 & 3.6
\end{array} \right)$  \\
\cline{2-3}
\multicolumn{1}{|c|}{} & $4$ & $\left(\begin{array}{cc}
13.4 & -8.9 \\ -8.9 & 13.4
\end{array} \right)$  \\
\cline{2-3}
\multicolumn{1}{|c|}{} & $5$ & $\left(\begin{array}{cc}
8.9 & -4.7 \\ -4.7 & 8.9
\end{array} \right)$\\
\cline{2-3}
\multicolumn{1}{|c|}{\multirow{-6}{*}[3em]{\begin{sideways}Interference\end{sideways}}} & $6$ & $\left(\begin{array}{cc}
11.6 & -3.7 \\ -3.7 & 11.6
\end{array} \right)$ \\
\cline{1-3}
\end{tabular}
\label{tab:as1}
\end{table}


\begin{IEEEbiography}[{\includegraphics[width=1in,height=1.25in,clip,keepaspectratio]{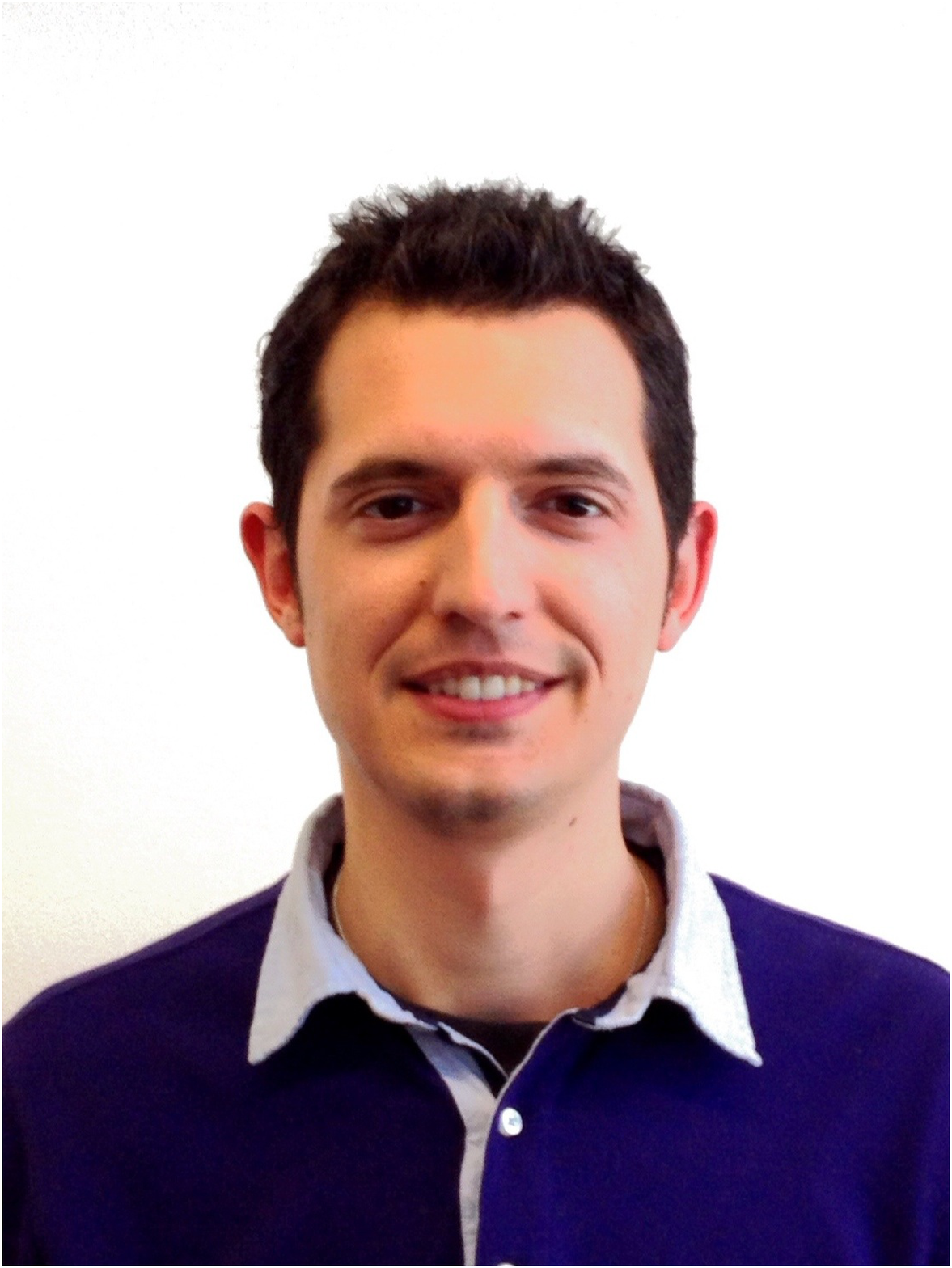}}]{Pol Henarejos}
was worn in Barcelona, Catalonia. He received the M.Sc. degree in telecommunication engineering from UPC in May 2009 and the European Master of Research on Information and Communication Technologies (MERIT) in 2012. 

He joined CTTC in January 2010 as a research engineer. In 2010, he participated in European projects implementing real receivers using the filterbank multicarrier approach. After that, he was involved in industrial projects based on implementations of the physical layer aspects of LTE in prototypes. That research permitted the deployment of the new standard for interactive services in satellite communications and testing innovative techniques. 

His interests comprise the implementations of the physical layer of radio communications in real devices to theoretical studies in resource management in multiuser scenarios.
\end{IEEEbiography}

\begin{IEEEbiography}[{\includegraphics[width=1in,height=1.25in,clip,keepaspectratio]{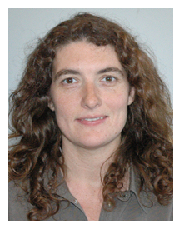}}]{Ana I. P\'{e}rez-Neira}
is full professor at UPC (Technical University of Catalonia) in the Signal Theory and Communication department.  

She has been the leader of 20 projects and has participated in over 50 (10 for ESA). She is author of 50 journal papers (20 related with Satcom) and more than 200 conference papers (20 invited). She is co-author of 4 books and 5 patents (1 on satcom). Since 2008 she is member of EURASIP BoD (European Signal Processing Association) and since 2010 of IEEE SPTM (Signal Processing Theory and Methods). She has been guest editor in 5 special issues and currently she is editor of IEEE Transactions on Signal Processing and of Eurasip Signal Processing and Advances in Signal Processing. She has been the general chairman of IWCLD’09, EUSIPCO’11, EW’14 and IWSCS’14. She has participated in the organization of ESA conference 1996, SAM’04 and she is the general chair of ASMS’16. She has been in the board of directors of ETSETB (Telecom Barcelona) from 2000-03 and Vicepresident for Research at UPC (2010-13). She created UPC Doctoral School (2011). Currently, she is Scientific Coordinator at CTTC (Centre Tecnològic de Telecomunicacions de Catalunya). She is the coordinator of the Network of Excellence on satellite communications, financed by the European Space Agency: SatnexIV.

Her research topics are in: multi-antenna signal processing for satellite communications and wireless and in physical layer scheduling for multicarrier systems.
\end{IEEEbiography}

\end{document}